\documentclass[a4paper,11pt]{article}
\usepackage{jinstpub} % for details on the use of the package, please see the JINST-author-manual
\usepackage{kotex}
\usepackage{array}

\title{Radiopurity measurements of liquid scintillator for the COSINE-100 Upgrade}
\author[a]{J.~Kim}
\author[a,1]{C.~Ha\note{Corresponding author.}}
\author[b]{S.H.~Kim}
\author[c,b]{W.K.~Kim}
\author[b,c]{Y.D.~Kim}
\author[d,2]{Y.J.~Ko\note{Corresponding author.}}
\author[b]{E.K.~Lee}
\author[c,b]{H.~Lee}
\author[b,c]{H.S.~Lee}
\author[b]{I.S.~Lee}
\author[b]{J.~Lee}
\author[b,c]{S.H.~Lee}
\author[e]{S.M.~Lee}
\author[a]{Y.J.~Lee}
\author[b]{G.H.~Yu}

\affiliation[a]{Department of Physics, Chung-Ang University,\\Seoul 06974, Republic of Korea}
\affiliation[b]{Center for Underground Physics, Institute for Basic Science(IBS),\\Daejeon 34126, Republic of Korea}
\affiliation[c]{IBS School, University of Science and Technology (UST), \\Daejeon 34113, Republic of Korea}
\affiliation[d]{Department of Physics, Jeju National University, \\Jeju 63243, Republic of Korea}
\affiliation[e]{Department of physics, Carnegie Mellon University, \\Pittsburgh, PA 15213, U.S.A.}

\emailAdd{chha@cau.ac.kr}
\emailAdd{yjkophys@jejunu.ac.kr}

\abstract{
A new 2,400\,L liquid scintillator has been produced for the COSINE-100 Upgrade, which is under construction at Yemilab for the next phase of the COSINE dark matter experiment phase. The linear-alkyl-benzene-based scintillator is designed to serve as a veto for NaI(Tl) crystal targets and a separate platform for rare event searches. We measured radioactivities of the liquid scintillator using a custom-made 445\,mL cylindrical Teflon container equipped with two 3-inch photomultiplier tubes. Analyses show activity levels of $0.091 \pm 0.042$\,mBq/kg for $^{238}$U and $0.012 \pm 0.007$\,mBq/kg for $^{232}$Th.
}

\begin{document}
\maketitle
\flushbottom

\section{Introduction}

Observational evidence for the existence of dark matter has been well-documented~\cite{DM_evidence_1,DM_evidence_2, Planck:2013mth}, leading to a surge in direct detection experiments~\cite{PhysRevLett.131.041002}. Among these, the DAMA experiment has claimed the observation of an annual modulation signal that aligns with the presence of weakly interacting massive particles~(WIMPs), a leading dark matter candidate~\cite{DAMA:2008jlt,Bernabei:2013xsa}. To independently test the DAMA experiment, the COSINE experiment was launched, using the same target material, NaI(Tl) crystals~\cite{Adhikari:2017esn}. After 6.5\,years of operation at the Yangyang Underground Laboratory, the COSINE-100 experiment has produced results challenging DAMA’s claims with a significance of more than 3\,standard deviations~\cite{Carlin:2024maf}. Preparations are now underway for the next phase, the COSINE-100 Upgrade~(COSINE-100U)~\cite{Lee:2024wzd}, at a new underground facility, Yemilab, in Jeongseon, Korea~\cite{Park:2024sio}.

A liquid scintillator~(LS) system will be used in the upcoming COSINE-100U experiment to help characterize signals and backgrounds in the crystal detectors. A similar system was used in the COSINE-100 experiment where the NaI(Tl) crystal assemblies are positioned at the center of an acrylic box, which is filled with LS, ensuring a minimum veto thickness of 40\,cm surrounding the crystal assembly at all points. Eighteen 5-inch photomultiplier tubes~(PMTs) attached to the box record light signals produced in the LS, and the data acquisition system simultaneously combines these signals with those from the crystals. Thus, the LS functions as a tagger for gamma rays originating both from outside the box and from the crystals themselves. The LS veto system has demonstrated tagging efficiencies of 15--30\,\% in the 2--6\,keV range, with 65--75\% efficiency for the detection of $^{40}$K originating from the crystals~\cite{Adhikari:2020asl}. Tagged events can also serve as an independent channel for understanding the background~\cite{COSINE-100:2024ola}. 
Additionally, the LS is utilized for exotic dark matter searches when the coincidence interactions simultaneously with crystals occur~\cite{Borexino:2012guz, COSINE-100:2018ged}.

A total of 2,400\,liters of the linear-alkyl-benzene~(LAB) based LS has been newly produced and is awaiting use in the COSINE-100U experiment. Three\,grams of 2,5-diphenyloxazole~(PPO) were added per liter of LAB to act as a fluorescent material.
Since the emission spectrum of PPO does not match the spectral response of the PMT, 30\,mg/L of 1,4-bis[2-methylstyryl]benzene~(bis-MSB) was added as a wavelength shifter for efficient light collection~\cite{PARK201345}. The radiopurity of the LAB-LS was measured prior to installation in the main detector to minimize the risk of introducing significant background radioactivity to the crystals. Data were collected for one month using a small-scale detector at a ground laboratory. Using the data, we developed pulse shape discrimination~(PSD) and $\beta/\alpha$ coincidence analysis to determine the intrinsic background level of the liquid. Furthermore, radioactivity measurements of the LAB-LS sample were performed using high purity germanium detectors~\cite{Sala:2016wlz} for comparison.

\section{Experimental setup for radiopurity measurements}

\begin{figure}[htbp]
\centering
\includegraphics[width=.9\textwidth]{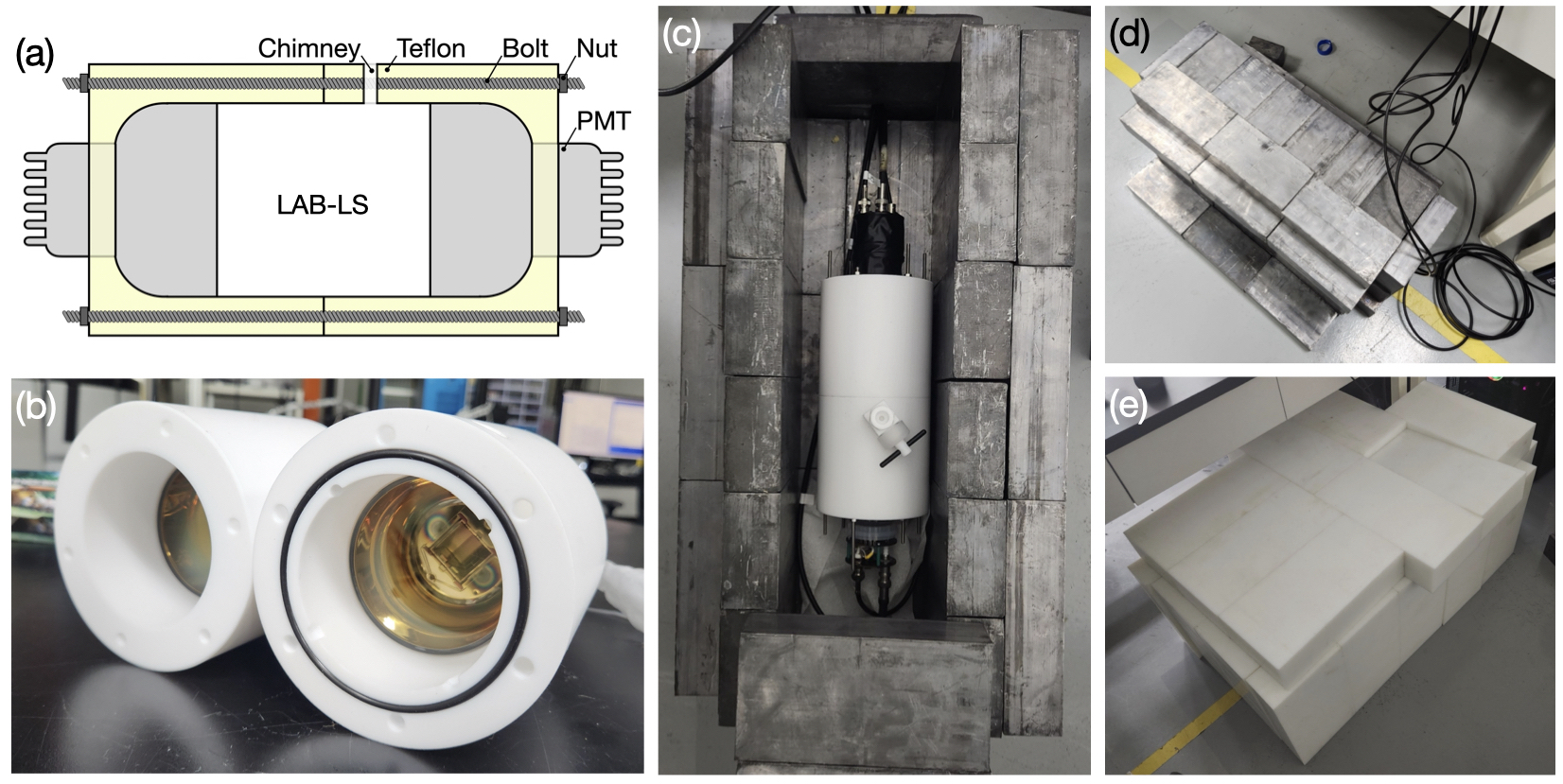} 
\caption{(a) A schematic diagram of the detector. The yellow region denotes the PTFE container and the gray region denotes the PMTs on both sides. (b) A cross-sectional view of the actual detector, showing the teflon housing and PMT window. (c) The detector assembly inside the lead castle. (d) An external view of the lead shielding before installation of polyethylene blocks. (e) An external view of the polyethylene block shielding structure. \label{fig:detector}}
\end{figure}

To quickly screen the LAB-LS for its radiopurity, a mini-detector designed to contain a 445\,mL sample of the liquid sample was built. As shown in Fig.~\ref{fig:detector}~(a), two cylindrical-hollow polytetrafluoroethylene (PTFE) cases with embedded Hamamatsu R12669SEL\footnote{selected for high quantum efficiency} PMT are designed so that the space between them can be filled with the liquid. The inner radius of the case matches the radius of the PMT to mechanically stabilize the PMTs and prevent leakage through the gap between the case and PMT. The outer radius of the case is 7.6\,cm, and there are six 3\,mm radius holes that go through the cases as shown in Fig.~\ref{fig:detector}~(a) and (b). 
When the two PTFE cases are joined, an O-ring is placed at the matching area, and long bolts are inserted into the six holes, with nuts tightened on the outside of the cases. After filling the LAB-LS through the chimney, the chimney is sealed with a PTFE stopper, as shown in Fig.~\ref{fig:detector}~(c), to prevent contamination by external radioactivities, most notably radons.

The assembled detector was shielded with lead bricks to reduce external background, as shown in Fig.~\ref{fig:detector}~(c) and (d). An additional layer of polyethylene blocks was added for environmental neutron mitigation (see Fig.~\ref{fig:detector}~(e)). The analog signal pulses were digitized, and a full waveform is recorded using analog-to-digital converter~(ADC) modules with a 2\,ns sampling precision. An event was triggered only when both PMT signals exceeded 6\,mV discrimination threshold. With a trigger rate of 6\,Hz, a total of 522\,hours of data was collected over the course of a month. 

\section{Detector response}
\subsection{Particle identification}
\label{sec:psd}
\begin{figure}[htbp]
    \centering
    \begin{tabular}{cc}
    \includegraphics[width=.45\textwidth]{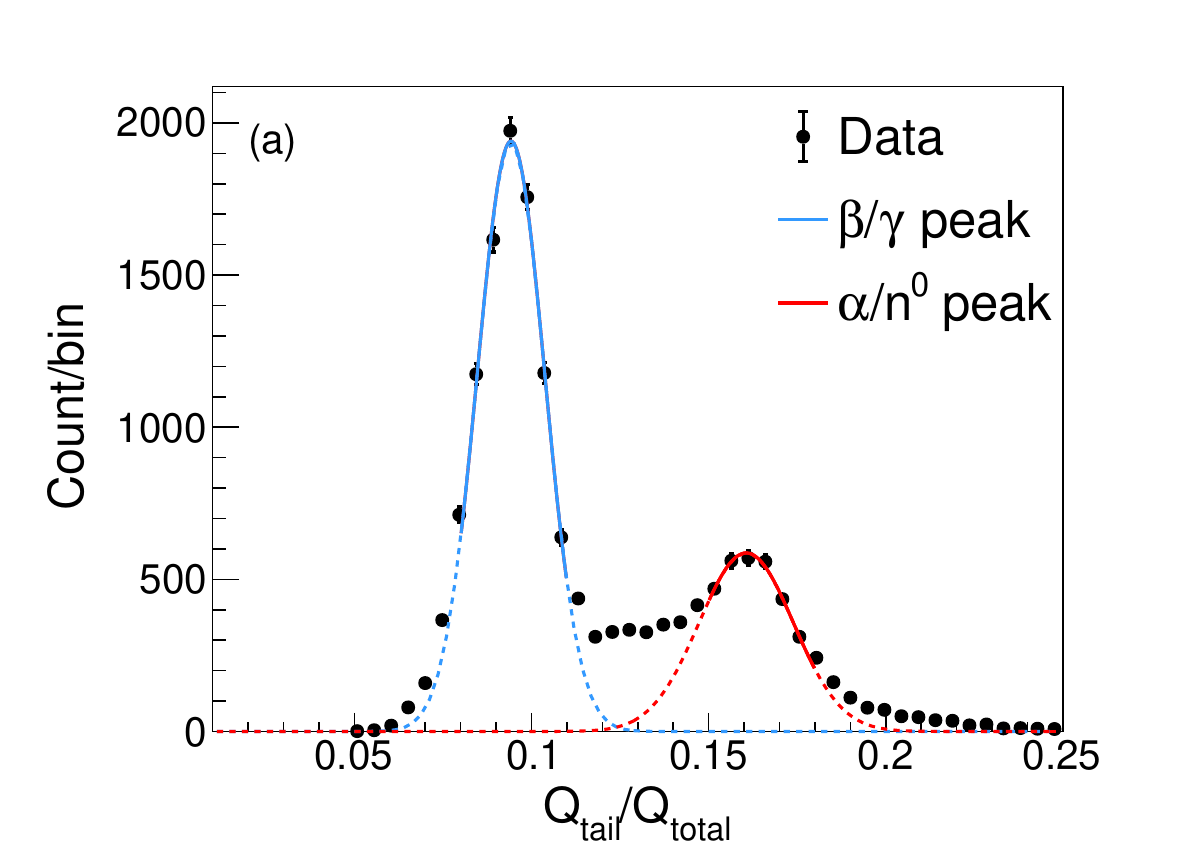} &
    \includegraphics[width=.45\textwidth]{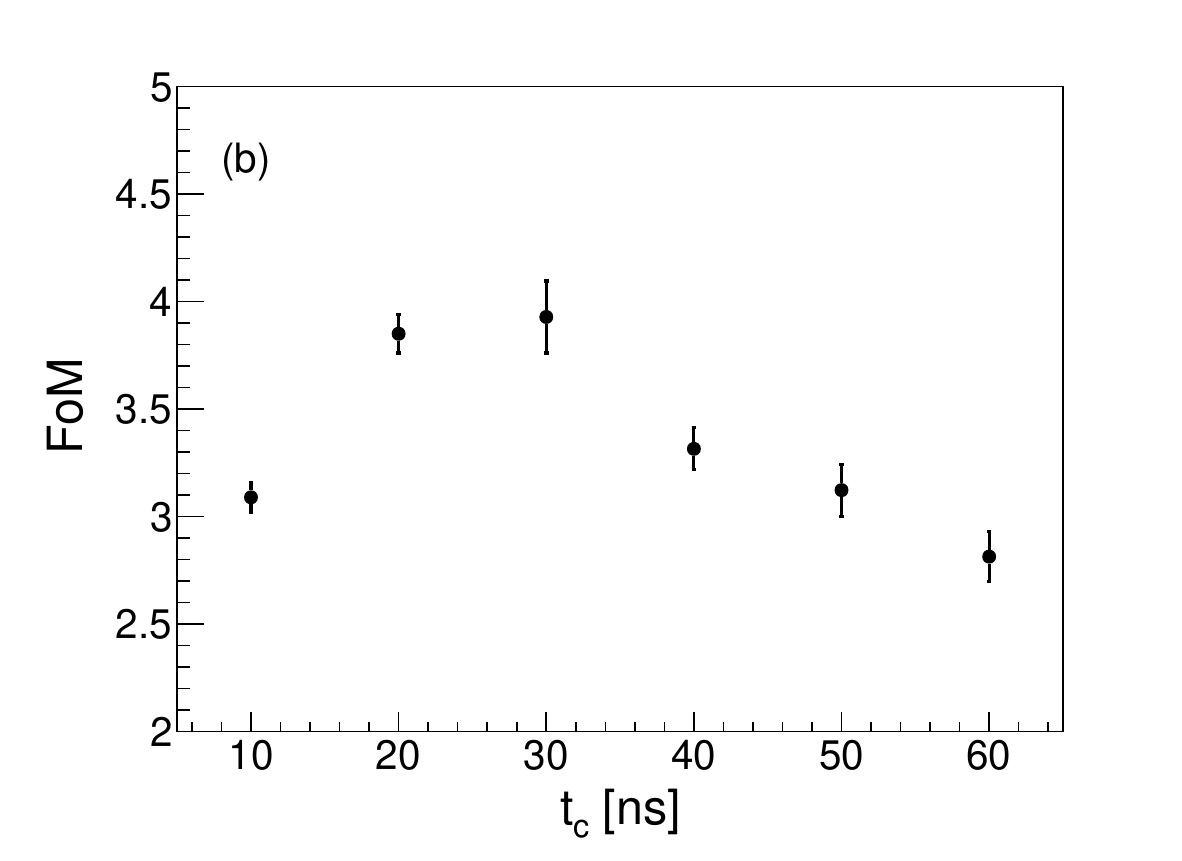} 
\end{tabular}
\caption{$Q_\mathrm{tail}$ optimization. As varying the $t_c$ from 10\,ns to 60\,ns, draw the PSD parameter distribution as (a) and fit it with Gaussian function(blue and red curve). The FoM representing the separation between the $\beta/\gamma$ and $\alpha/n^0$ was derived from these fits, as shown on (b). The  FoM peaked at 3.9 when $t_c$ is 30\,ns.}
\label{fig:psdA}
\end{figure}

Pulse shape discrimination~(PSD), which distinguishes between $\beta/\gamma$ and $\alpha/n^0$ based on the decay time of the liquid, is a popular technique for particle identification~(PID). To identify the difference in decay time from a waveform, a PSD parameter is defined as the ratio of $Q_\mathrm{tail}$ to $Q_\mathrm{total}$,
\begin{equation}
    \frac{Q_\mathrm{tail}}{Q_\mathrm{total}}
      = \frac{\sum_{t_E-t_c}^{t_E}q_i}{\sum q_i},
\end{equation}
where $t_c$ is the range of summation for $Q_\mathrm{tail}$, while $t_E$ and $q_i$ represents the end time bin and the charge of $i^\mathrm{th}$ bin of the waveform, respectively. Figure~\ref{fig:psdA}~(a) shows the distribution of PSD parameter when $t_c$ is 30\,ns. One can see the $\beta/\gamma$ events are distributed from 0.07 to 0.12 while $\alpha/n^0$ events are distributed from 0.14 to 0.18. The decay time of $\beta/\gamma$ events is relatively short, and thus appear in the lower range of PSD parameter, while $\alpha/n^0$ is distributed in the higher range. The data regions are fitted with models based on Gaussian functions for the $\beta/\gamma$ and $\alpha/n^0$ separately. Using the resulting fit parameters, the discriminative power of PSD is quantified by the figure of merit~(FoM), which is defined as follows,
\begin{equation}
    \mathrm{FoM} = \frac{\mu_{\alpha/n^0} - \mu_{\beta/\gamma}}{ \sqrt{\sigma_{\alpha/n^0}^2 + \sigma_{\beta/\gamma}^2} },
    \label{eq:FoM}
\end{equation}
where $\mu$ and $\sigma$ are the mean and standard deviation of the best-fit, respectively. Since the FoM depends on the $t_c$, we optimize the FoM by varying the $t_c$ from 10 to 60\,ns. Figure~\ref{fig:psdA}~(b) shows the scan results of the FoM as a function of $t_c$. We determined $t_c$ to be 30\,ns with the largest FoM value.

\begin{figure}[htbp]
\centering
\includegraphics[width=0.9\textwidth]{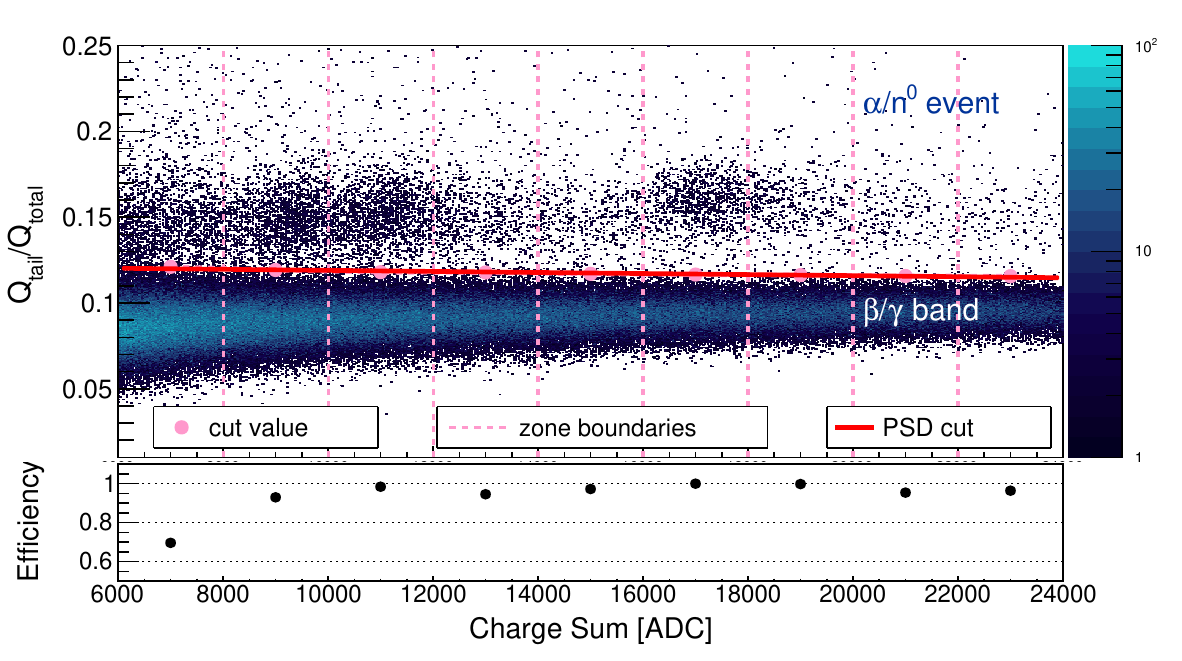}
\caption{The PSD result for distinguishing $\beta/\gamma$ and $\alpha/n^0$ events. The clear band below the red line shows the distribution of $\beta/\gamma$ events, while the band with three islands above the curve represent the distribution of $\alpha/n^0$ events. The red line was used to separate $\alpha/n^0$ events from $\beta/\gamma$ events.
\label{fig:psdB}}
\end{figure}

The charge-dependent distribution of the PSD parameter with the optimized $t_c$ can be seen in Fig.~\ref{fig:psdB}. One can see a clear distinction between the $\beta/\gamma$ and $\alpha/n^0$ bands separated by the red line. The three islands in the $\alpha/n^0$ band represent $\alpha$ events and a continuous underlying distribution for environmental $n^0$. To determine the criteria for $\alpha/n^0$ selection, the data was divided into nine zones based on a two-PMT total charge, from 6,000\,ADC to 24,000\,ADC. For each zone, the data on the PSD parameter was fitted with two separate Gaussian functions, and then we found a cut value that rejects 99.87\,\% of $\beta/\gamma$ events. These cut values are plotted in Fig.~\ref{fig:psdB} as pink dots, and a linear function was obtained by fitting them to use as a selection criterion. Here, the PSD cut shows an efficiency of over 94.51\,\% for $\alpha/n^0$ events over 10,000\,ADCs.

\subsection{Energy calibration}
\label{sec:calib}
\begin{figure}[htbp]
\centering
\begin{tabular}{ccc}
\includegraphics[width=0.36\textwidth]{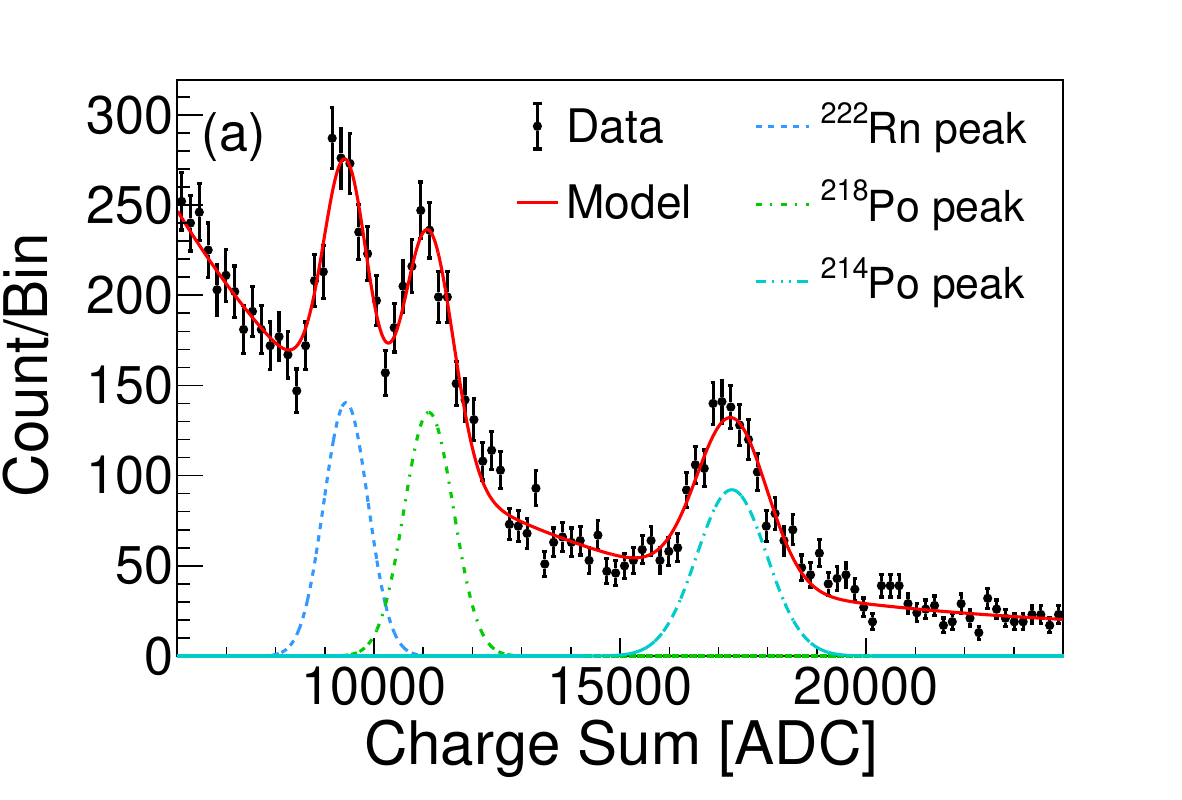} &
\includegraphics[width=0.27\textwidth]{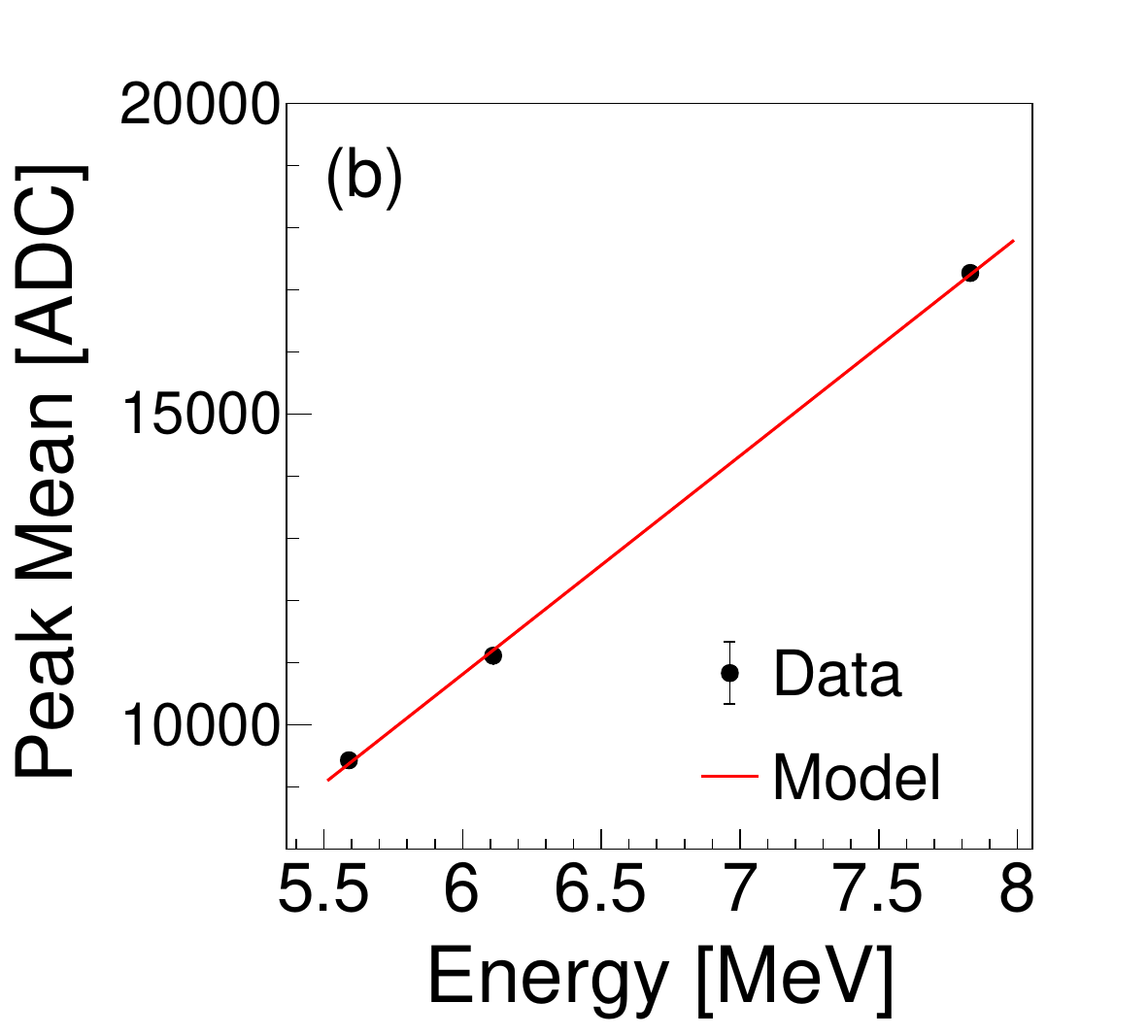} &
\includegraphics[width=0.27\textwidth]{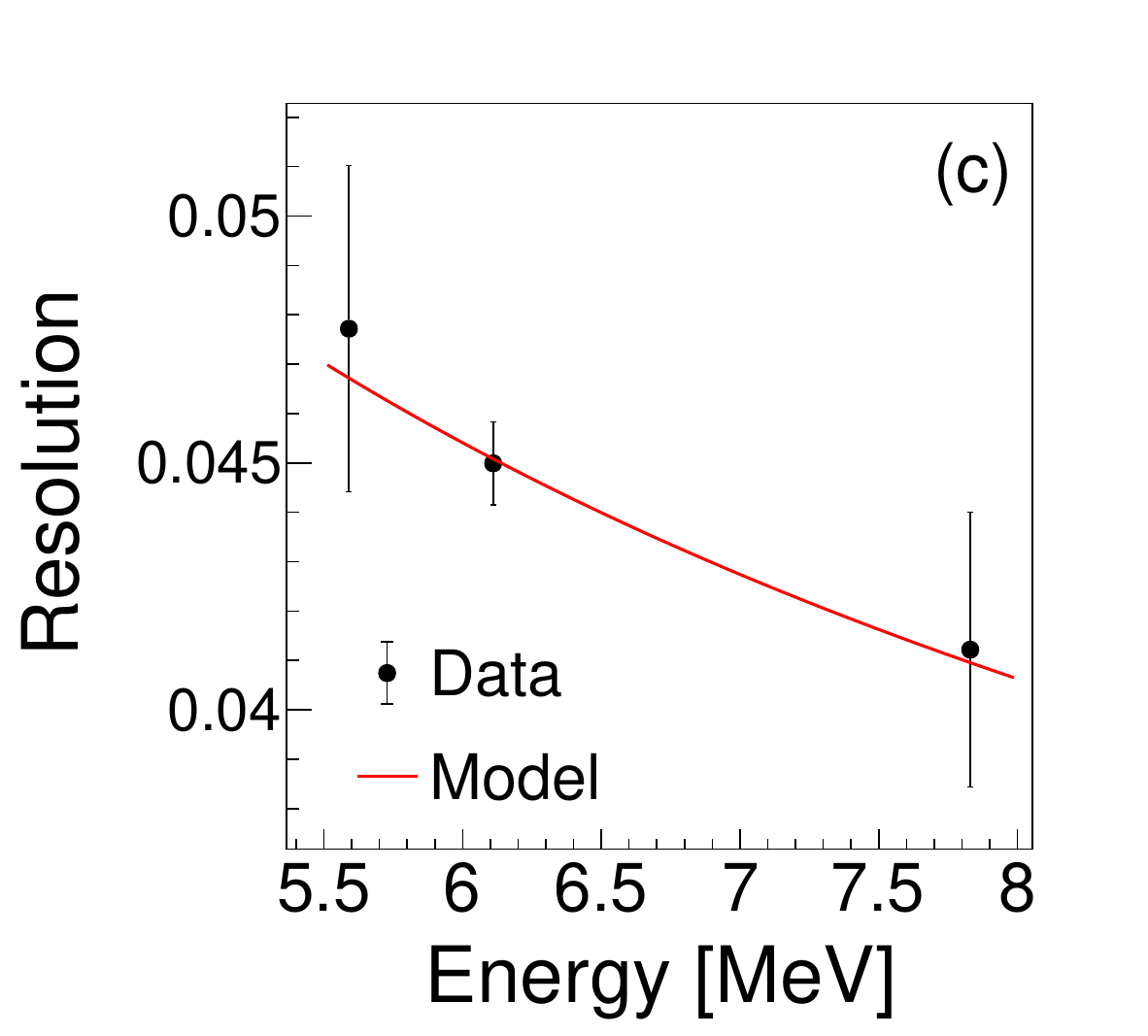}     
\end{tabular}
\caption{Energy Calibration. (a) Three Gaussian functions with an exponential and constant term for background were used to fit the $^{222}$Rn, $^{218}$Po and $^{214}$Po peaks. The red solid line indicates the fitted model, and the three Gaussian peaks are represented by the dashed lines. Total three means and three widths of each peaks were obtained and these are used for the energy calibration function (b) and the resolution function (c). \label{fig:calibration}
The black data points with error bars were obtained from the fitting results with known energies of three $\alpha$ peaks. 
}
\end{figure}

The three islands identified in Fig.~\ref{fig:psdB} appear as three distinct peaks in the charge distribution of the $\alpha/n^0$ events selected via PID in Sec.~\ref{sec:psd} as shown in Fig.~\ref{fig:calibration}~(a). They appear to come from $\alpha$ peaks caused by $^{222}$Rn contaminated during LS production and detector assembly, and the charge ratios between them are in good agreement with the ratios between the $\alpha$ decay energies of $^{222}$Rn, $^{218}$Po, and $^{214}$Po which allows for an $\alpha$ energy calibration.

The charge distribution of the $\alpha/n^0$ events was modeled with three Gaussian functions to describe the three $\alpha$ peaks and one exponential function with a constant term to account for the background consisting mainly of fast neutrons. The means and standard deviations of the Gaussian functions were used to determine the calibration and resolution functions for $\alpha$ energy. The Gaussian mean values of the charge sum distributions were fitted by a linear function and used as a calibration function for $\alpha$ energy, while the resolutions were fitted by,
\begin{equation}
    \frac{\sigma_E}{E} = \sqrt{\frac{A}{E} + B},
\end{equation}
where $\sigma_E$ is the energy resolution, while $A$ and $B$ are used as free parameters. The fitted functions for $\alpha$ energy calibration and resolution are shown as red solid lines in Fig.~\ref{fig:calibration}~(b) and (c), respectively.
\section{Coincidence analysis}
\subsection{$\beta-\alpha$ from $^{238}$U chain}
To measure the activity of the LAB-LS $^{238}$U-contaminated series, the decays of $^{214}$Bi and $^{214}$Po in the decay chain were investigated. The isotope $^{214}$Bi decays to $^{214}$Po via $\beta$-decay with a Q-value of 3.27\,MeV. $^{214}$Po subsequently undergoes $\alpha$-decay with a Q-value of 7.83\,MeV and a half-life of 164.3\,$\mu$s to produce $^{210}$Pb. Thanks to the short decay time of $^{214}$Po, these sequential decays can appear as a time-coincidence event in the current DAQ setup.

\begin{figure}[htbp]
\centering
\begin{tabular}{cc}
\includegraphics[width=0.45\textwidth]{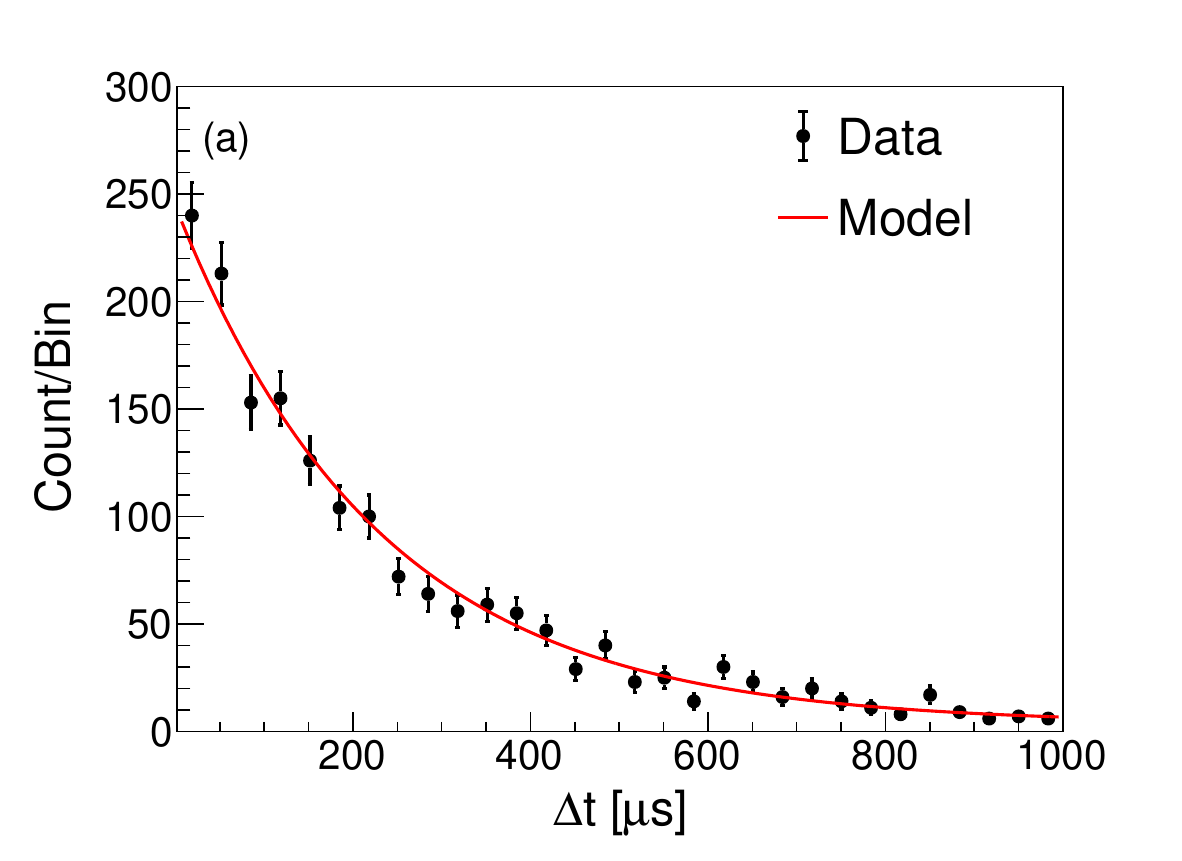} &
\includegraphics[width=0.45\textwidth]{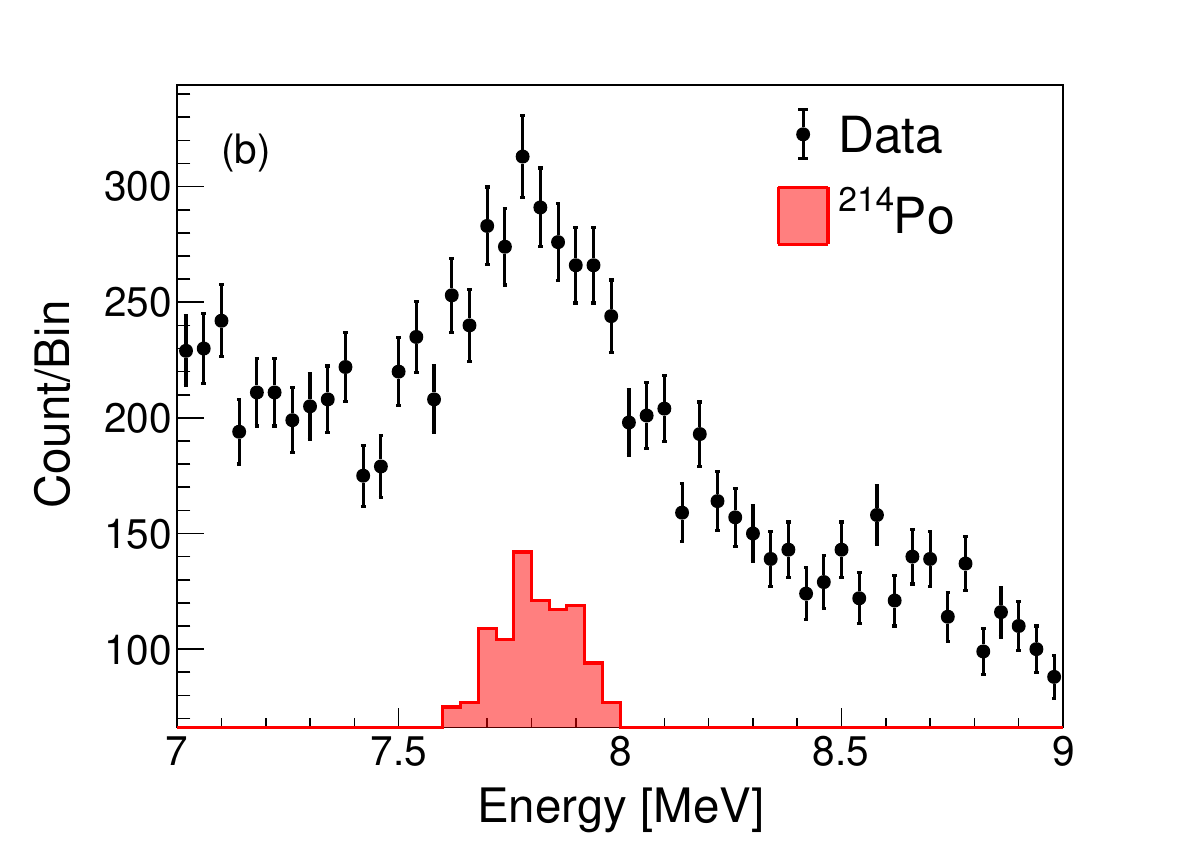} \\
\end{tabular}
\caption{Sequential decays of $^{214}$Bi and $^{214}$Po identified via $\beta-\alpha$ time coincidence. (a) Time difference $\Delta$t between $\beta$ and $\alpha$ events is fitted with an exponential function (red curve) to tag the sequential decays of $^{214}$Bi-$^{214}$Po~(black dots with error bars). (b) The energy distribution of $\alpha/n^0$ identified by the PSD method in Sec.~\ref{sec:psd}. The black dots denotes total $\alpha/n^0$ events, while the red-filled histogram represents the $\alpha$ events from $^{214}$Po decay.
\label{fig:Po214Tag}}
\end{figure}

Utilizing the PID from Sec.~\ref{sec:psd}, the time interval between $\beta$ and $\alpha$ over a range of 2 to 1000\,$\mu$s is shown in Fig.~\ref{fig:Po214Tag}~(a). When $\beta-\alpha$ time difference is within this time span, the energy distribution of those coincident $\alpha$ events is represented by the tagged events in Fig.~\ref{fig:Po214Tag}~(b). The coincidence time between $\beta$ and $\alpha$ events can be determined by fitting the data with an exponential function. Since random coincidences between $\beta/\gamma$ and $\alpha/n^0$ events may occur, we used an exponential function with an additional constant term,
\begin{equation}
    f(\Delta t) = A\,\exp\left(-\frac{\Delta t}{t_{1/2}/\ln2}\right) + C,
    \label{eq:fitfunc}
\end{equation}
where the variable $\Delta t$ is coincident time difference. The three fitting parameters are $A$, $t_{1/2}$, and $C$. Here, $A$ represents the activity of $^{210}$Po, $t_{1/2}$ is its half-life, and $C$ accounts for the amount of random coincidence. We obtained the best-fit half-life of $t_{1/2}=160.1 \pm 5.1$\,$\mu$s, which is consistent with the known value.

\begin{figure}[htbp]
\centering
\includegraphics[width=0.9\textwidth]{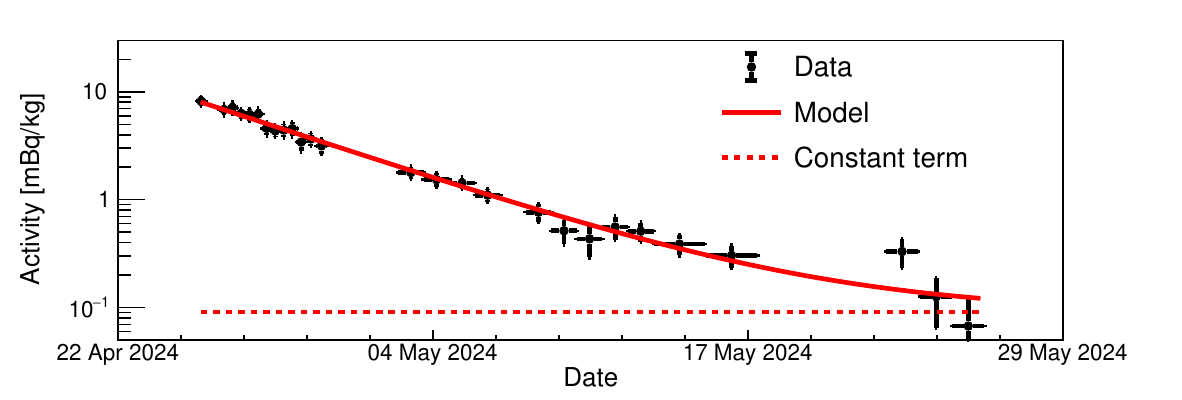}
\caption{Time evolution of the measured activity. The black dots with error bars represent the activity (in mBq/kg) at each time interval. They were fitted with a function represented by the red solid line, of which the constant term is depicted by the red dashed line.
\label{fig:Po214Activity}}
\end{figure}

A significant portion of the remaining $\beta-\alpha$ time-coincidence events is attributed to contamination from $^{222}$Rn (with a half-life of 3.8\,days). Consequently, it is necessary to analyze these events over an extended period. The collected data over a month were divided into appropriate time intervals for this analysis. The $\Delta t$ distribution in each interval was fitted by Eq.~\ref{eq:fitfunc} with the half-life fixed at 164.3\,$\mu$s, while keeping $A$ and $C$ as free parameters.

The activity $A$, determined from the fitting, is plotted in Fig.~\ref{fig:Po214Activity}. The observed decrease in activity appears to be due to contamination from $^{222}$Rn prior to detector assembly. To model this decrease, we used the same function as in Eq.~\ref{eq:fitfunc}. The exponential term represents the contribution from $^{222}$Rn, while the constant term accounts for internal contamination from the $^{238}$U series. The best-fit model aligns well with the observed decrease in activity. The internal contamination from $^{238}$U, as determined from the constant term, is $0.091 \pm 0.042$\,mBq/kg. 

\subsection{$\alpha-\alpha$ from $^{232}$Th chain}
The sequential $\alpha$\,decays of $^{220}$Rn and $^{216}$Po in the $^{232}$Th decay chain can be identified by time-coincidence condition, thanks to the short $^{216}$Po's half-life ($t_{1/2}[^{216}\rm Po]=140\,ms$). A dataset consisting of $\alpha/n^0$-only events identified via PID was prepared, and the $\alpha-\alpha$ coincidence events were selected using two criteria: We require the time interval between two consecutive $\alpha$ events to be less than 1\,second, with energy selection within 5 standard deviations of the measured $\alpha$ energies. The standard deviations are based on detector resolutions modeled using Gaussian functions, with the known $\alpha$ energies as the means.

\begin{figure}[t]
\centering
\includegraphics[width=0.9\textwidth]{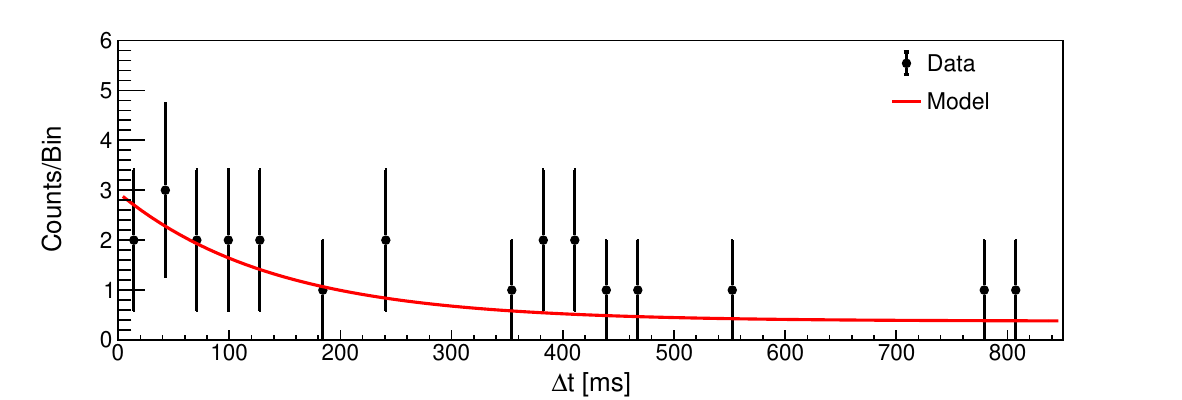} 
\caption{The $\alpha-\alpha$ coincidence events by a sequential decays of $^{220}$Rn and $^{216}$Po. The black dots with error bars denote the measured time intervals $\Delta$t between coincident $\alpha$ events, while the red line represents the exponential fit based on Eq.~\ref{eq:fitfunc}. 
}
\label{fig:Th232}
\end{figure}

Similarly, through $\alpha-\alpha$ coincidence analysis, we obtained an activity of $0.012 \pm 0.007$\,mBq/kg. In Fig.~\ref{fig:Th232}, the time difference of selected $\alpha-\alpha$ coincidence events are represented as data points with error bars. 
The coincidence time distribution was fitted by Eq.~\ref{eq:fitfunc}, with the half-life fixed at 140\,ms. The contamination by $^{232}$Th is derived from the amplitude term of the best-fit model.
\section{Summary and discussion}

In the COSINE dark matter search project, the unique LS veto system plays a crucial role in background reduction and detector understanding. The next phase of COSINE-100, named as COSINE-100U, is under preparation, and the total of 2,400\,L of LAB-based LS has been produced for use in the veto system. As for the performance testing of the newly produced LAB-LS, the activities of $^{238}$U and $^{232}$Th are measured.

A small-sized radiopurity screening detector is created and one month of data was collected for the test. With the PSD technique, energy calibration of $\alpha/n^0$ events is performed by separating $\alpha/n^0$ events from the $\beta/\gamma$ sample. The $\beta-\alpha$ coincidence analysis of sequential decays of $^{214}$Bi and $^{214}$Po in the $^{238}$U decay chain provides measurements of internal contamination by $^{238}$U and additional contamination by $^{222}$Rn, resulting in a measured $^{238}$U activity of $0.091 \pm 0.042$\,mBq/kg. Additionally, the sequential $\alpha$ decays of $^{222}$Rn and $^{216}$Po in the $^{232}$Th decay chain were identified, determining the $^{232}$Th activity to be $0.012 \pm 0.007$\,mBq/kg. These results are consistent with those of the liquid scintillator used in the COSINE-100 experiment, which reported activities of 0.087\,mBq/kg and 0.016\,mBq/kg for $^{238}$U and $^{232}$Th, respectively~\cite{Adhikari:2020asl}.

\begin{figure}[htbp]
\centering
\begin{tabular}{cc}
\includegraphics[width=0.45\textwidth]{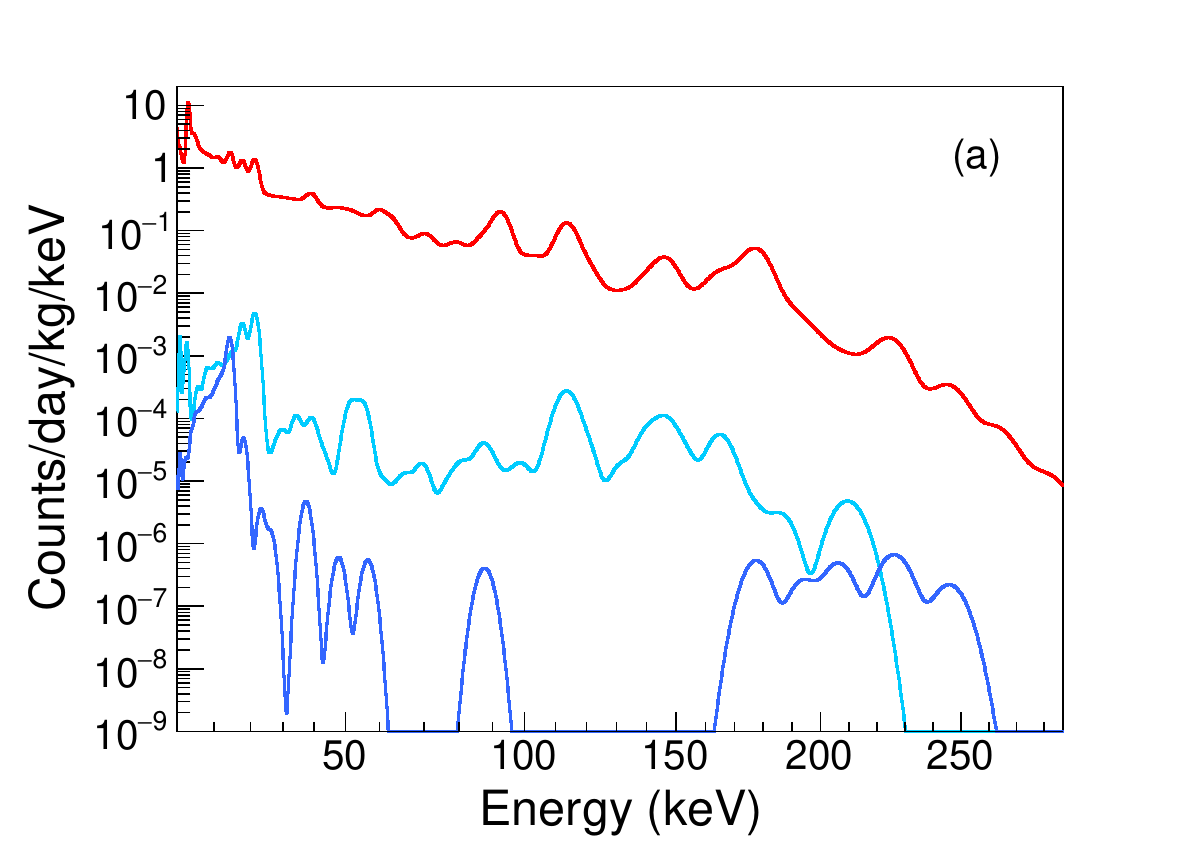} &
\includegraphics[width=0.45\textwidth]{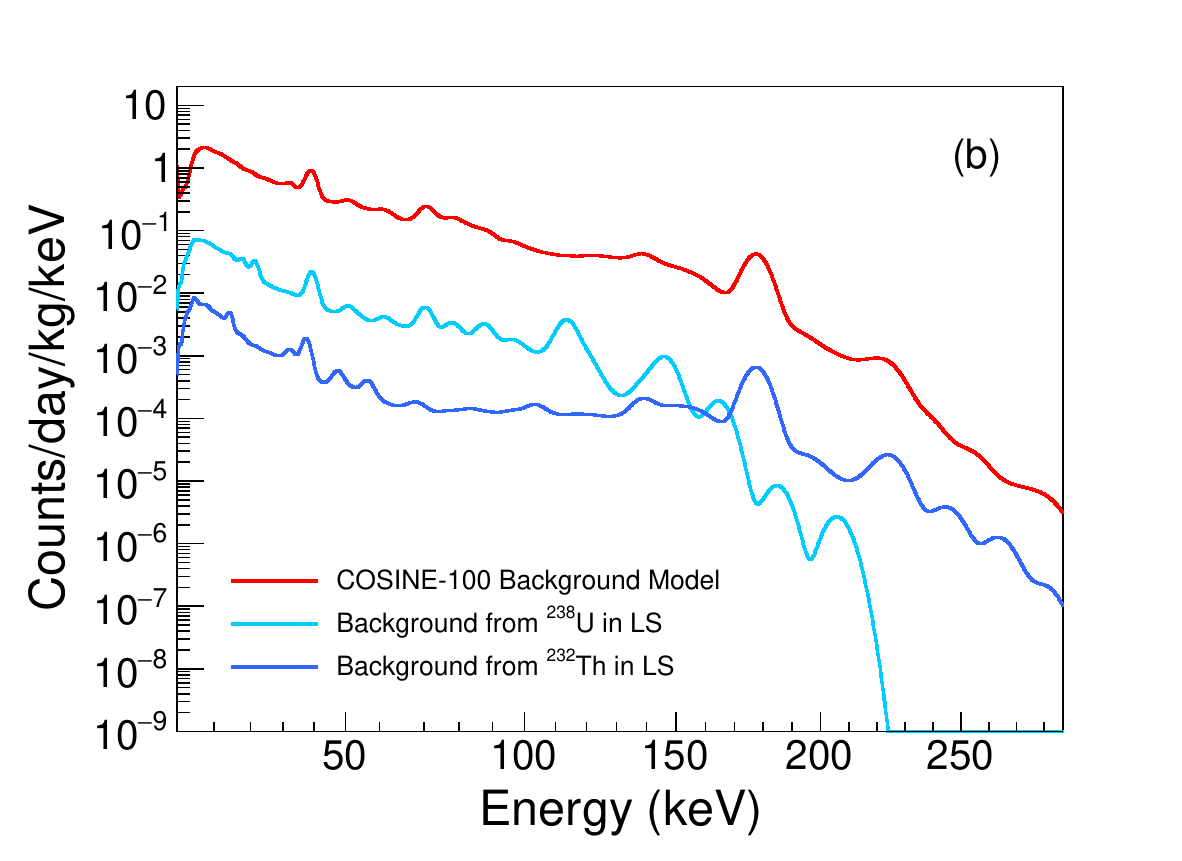} \\
\end{tabular}
\caption{Simulated background model. The red histogram indicates the total COSINE-100 background model and the cyan and blue histograms indicate the contributions from $^{238}$U and $^{232}$Th contamination in LS, respectively. The single-hit energy spectrum (a) shows contributions of 0.07\,\%, while the multiple-hit spectrum (b) indicates contributions of 3.14\,\%. These low levels illustrate the minimal impact of LS contamination compared to the total COSINE-100 background model.
\label{fig:BkgdModel}}
\end{figure}

In order to check the influence of the measured activities on the background of the NaI(Tl) crystal, we compared those with the background model of COSINE-100~\cite{COSINE-100:2024ola}. In Fig.~\ref{fig:BkgdModel}, one can see that the overall contributions inferred from the measured results are negligible for the dark matter search region and, in particular, the contribution at the region of interest for WIMP signals (1-6\,keV of single-sited interaction region) is lower than 0.02\,\%.
Hence, we confirmed that the LAB-LS sample is sufficiently pure to be used as the veto system in the COSINE-100U experiment.

\paragraph{Activity measurements using HPGe detector}
For cross-validation, the radioactivities of the LAB-LS sample were also measured using a High Purity Germanium~(HPGe) detector~\cite{Lee:2019yICHEP}. The HPGe detector collected data for 17.25\,days in Yemilab. The results for four radionuclides are presented in Table~\ref{tab:HPGeResult}, showing that all measurements were within the upper limits, indicating that the LAB-LS is sufficiently clean. 

\begin{table}[htbp]
\centering
\caption{Measured radioactivity levels of the LAB-LS sample using an HPGe detector.\label{tab:HPGeResult}}
\smallskip
\begin{tabular}{c|cccc}
\hline
Nucleus & $^{238}$U & $^{40}$K & $^{228}$Ac & $^{228}$Th \\
\hline
Activity [mBq/kg] & < 0.90 & < 7.39 & < 2.01 & < 0.74 \\
\hline
\end{tabular}
\end{table}

%\appendix
%\section{Appendix Example}
%Please always give a title also for appendices.

\acknowledgments
%We thank the Institute for Basic Science (IBS) Research Solution Center (RSC) for providing high performance computing resources. 
This research was supported by: 
the Institute for Basic Science under project code IBS-R016-A1 Republic of Korea,
the National Research Foundation of Korea (NRF) grant funded by the Korean government (MSIT) (NRF-2021R1A2C1013761),
and the Chung-Ang University Research Scholarship Grants in 2023. 

\bibliographystyle{JHEP}
\bibliography{biblio.bib}

\end{document}